\documentclass[lettersize,journal]{IEEEtran}
\usepackage{graphicx}
\usepackage{algorithm}
\usepackage{algorithmic}
\usepackage{adjustbox}  % 导言区
\usepackage{graphicx}   % 确保加载 graphicx 包
\usepackage{caption}
\usepackage{float}      
\usepackage{amsmath,amssymb,amsfonts}
\usepackage{subcaption}  % 在导言区添加
\usepackage{upgreek}
\usepackage{stfloats}
\usepackage{amsmath,amsfonts}
\usepackage{algorithmic}
\usepackage{algorithm}
\usepackage{array}
\usepackage[caption=false,font=normalsize,labelfont=sf,textfont=sf]{subfig}
\usepackage{textcomp}
\usepackage{stfloats}
\graphicspath{{images/}}
\usepackage{url}
\usepackage{verbatim}
\usepackage{graphicx}
\usepackage{cite}
\usepackage[normalem]{ulem}
\usepackage{bm}
\usepackage{amsmath}
\usepackage{multirow}
\usepackage{float}         % 图片浮动位置
\usepackage{overpic} 
\usepackage{bm}

\usepackage{color}
\usepackage{xurl}
\usepackage{makecell}

\usepackage{amsthm}
\IEEEoverridecommandlockouts

\usepackage[a4paper, total={184mm, 169mm}]{geometry}
\def\BibTeX{{\rm B \kern-.05em{\sc i\kern-.025em b} \kern-.08em
T\kern-.1667em\lower.7ex\hbox{E}\kern-.125emX}}
\ifCLASSOPTIONcompsoc

  \usepackage[nocompress]{cite}
\else
  \usepackage{cite}
\fi

\ifCLASSINFOpdf

\else
\fi

\begin{document}
\title{Mixed-Criticality Flow Scheduling with Low Delay and Limited Bandwidth in TSN}

\author{
\author{Wenyan Yan, Sijing Duan, \IEEEmembership{Member, IEEE,}
	and Dongsheng Wei, \IEEEmembership{Member, IEEE}
	\thanks{Wenyan Yan is with the School of Computer Science, Hunan First Normal University, Changsha, China. Sijing Duan is with the Department of Computer Science and Technology, Tsinghua University, Beijing, 100084, China. Dongsheng Wei is with the Key Laboratory for Embedded and Cyber-Physical Systems of Hunan Province, College of Computer Science and Electronic Engineering, Hunan University, Changsha, China. 
 	(E-mail: ywy33@hnu.edu.cn, duansj@tsinghua.edu.cn, and weidongsheng@hnu.edu.cn) (Corresponding author: Sijing Duan)}}.} 
\maketitle

\begin{abstract}
Time-Sensitive Networking (TSN) is a promising Ethernet protocol with time determinism, widely used in time-critical systems such as industrial automation, automotive networks, and avionics. By allocating dedicated time windows for time-sensitive flows, TSN enables deterministic transmission; however, as network traffic grows, multiple flows may contend for the same window, causing large delays. Frame aggregation can mitigate this by combining multiple small frames into a larger one, thereby reducing the number of frames and required time windows, but existing approaches typically handle only single-priority traffic and cannot fully utilize pre-allocated time windows. To address this limitation, we propose MCFS-2L, a mixed-criticality flow scheduling scheme with low delay and limited bandwidth usage. MCFS-2L first aggregates critical and non-critical frames with the same source and destination nodes and harmonic periods into a single frame, and then applies a dynamic reassembly and scheduling method that selectively disaggregates non-critical frames from unschedulable aggregated frames. Experimental results show that MCFS-2L increases the acceptance ratio of critical and non-critical flows by up to 4.78\% and 8.58\%, respectively, while reducing bandwidth utilization by up to 11.88\%.

\end{abstract}

\IEEEpeerreviewmaketitle

\section{Introduction}
\begin{figure}[t]
	\centering
	\includegraphics[width=0.45\textwidth]{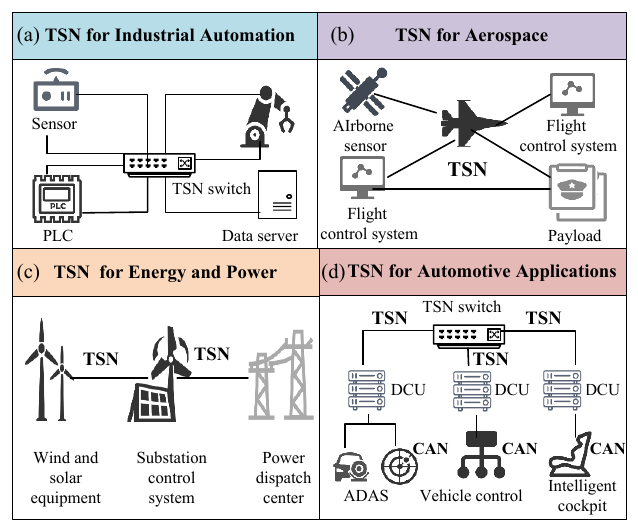}
	\caption{Applications of TSN.}
	\label{TSN network topology}
\end{figure}

Time-Sensitive Networking (TSN) is widely used in time-critical systems such as industrial automation, aerospace, energy and power, and automotive networks because of its high bandwidth and real-time capabilities \cite{wei2025refinedts}, as shown in Fig. \ref{TSN network topology}. TSN achieves deterministic data transmission by allocating a dedicated time window for each time-sensitive flow (defined by IEEE 802.1Qbv \cite{Qbv}) \cite{xue2025survey, long2025fastscheduler}. As a result, TSN is increasingly adopted as the backbone network for inter-domain communication in domain-centralized automotive architectures. Fig. \ref{TSN network topology}(d) illustrates a representative domain-centralized automotive architecture, which consists of three main domains: Advanced Driver Assistance Systems (ADAS), the vehicle control domain, and the intelligent cockpit domain \cite{bandur2021making, yan2025pfv2}. TSN is used for intra-domain communication in the ADAS and intelligent cockpit domains, while the vehicle control domain employs the Controller Area Network (CAN) bus (with low cost and high reliability) for internal communication. The TSN switch serves as a bridge, connecting the Domain Control Units (DCUs) across different domains.

As network traffic greatly increases due to intelligent applications such as ADAS \cite{bello2018recent}, the multiple data flows may compete for the same
time window and thus lead to large transmission delays.
To cope with the growing challenge of high traffic loads, frame aggregation has been introduced into this domain-centralized automotive architecture\cite{xie2023high}, \cite{wu2024real}. Frame aggregation combines multiple small data frames into a single larger frame to reduce the number of frames and time windows. The authors of Refs. \cite{xie2023high} and \cite{yan2024conflict} proposed frame aggregation solutions by aggregating multiple CAN frames into a TSN frame in the DCU to improve the schedulability. However, 
existing solutions often only focus on frames with a
single priority and fail to fully utilize the pre-allocated time windows in TSN. Therefore, there is an urgent need for more advanced frame aggregation strategies that can effectively process multi-priority frames and make full use of the available time windows in TSN.

In this study, we consider the multi-priority frames during frame aggregation, but face the following challenges. In TSN, different types of flows have their respective criticality requirements. Critical flows, such as safety data or real-time control information, need to be transmitted within strict time windows, while non-critical flows can tolerate a certain delay. Therefore, frame aggregation must ensure that critical traffic is delivered on time to meet its stringent timeliness requirements. Moreover, latency and bandwidth overhead are two key factors of flow transmission in TSN. The critical flow requires deterministic transmission, whereas non-critical flow typically aims to optimize bandwidth usage without strict timing constraints. Although frame aggregation reduces bandwidth overhead by merging multiple small frames into a larger one, it also increases the transmission delay of the aggregated frame. Therefore, an effective trade-off must be achieved between latency and bandwidth overhead.

To address these challenges, we develop MCFS-2L, a mixed-criticality flow scheduling technique that achieves low delay with limited bandwidth usage. MCFS-2L first performs mixed-criticality frame aggregation by merging critical and non-critical frames with the same source and destination nodes and harmonic periods into a single frame, thereby reducing the number of transmitted frames, the required time windows in the Gate Control List (GCL), and the bandwidth overhead of mixed-criticality traffic. It then applies a dynamic reassembly and scheduling method: when an aggregated frame becomes unschedulable, the non-critical frames within it are dynamically identified and removed one by one until the remaining frames form a schedulable frame; the extracted non-critical frames are subsequently re-aggregated into a new frame for rescheduling. The main contributions are as follows.

\begin{itemize}
\item We introduce MCFS-2L, a mixed-criticality flow scheduling framework tailored to domain-centralized automotive architectures under limited bandwidth.  
\item We design an integrated mixed-criticality frame aggregation and dynamic reassembly mechanism that jointly reduces frame count and GCL time windows while adaptively disaggregating non-critical traffic from unschedulable aggregated frames to enhance the acceptance ratio.  
\item We conduct extensive experiments on realistic automotive workloads to evaluate MCFS-2L, demonstrating higher acceptance ratios for both critical and non-critical flows and lower bandwidth utilization compared with state-of-the-art methods.
\end{itemize}

\section{Models for TSN}
This section describes the network model, TSN switch model, and flow model in detail.

\subsection{Network Model}
\label{section:Network model}
\begin{figure}[ht]
	\centering
\includegraphics[width=0.40\textwidth]{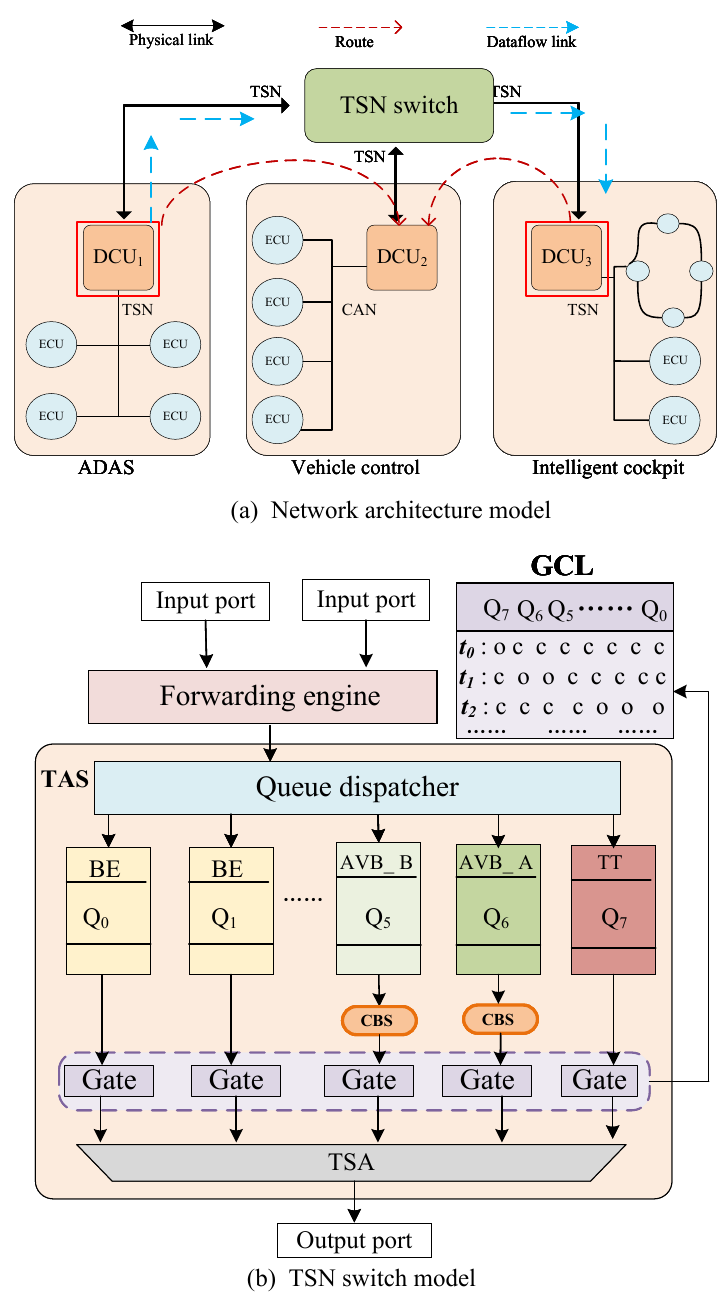}
	\caption{Model for TSN.}
	\label{TSN switch model}
\end{figure}

We take Fig. \ref{TSN switch model}(a) as an example to illustrate the network architecture model. The $\rm {DCU_1}$, $\rm {DCU_2}$, and $\rm {DCU_3}$ represent the Domain Control Unit of ADAS, vehicle control, and intelligent cockpit domains, respectively. The TSN switch in Fig. \ref{TSN switch model}(a) is responsible for connecting the different DCUs by full-duplex physical links. In Fig. \ref{TSN switch model}(a), the solid black line with a bidirectional arrow denotes physical links; the blue dotted line with a one-way arrow denotes the dataflow link; the dark red dotted line denotes the route. We represent the network architecture model as a directed graph, where all domain control units and the switch constitute the set of nodes, and all physical links constitute the set of edges. The route represents the transmission path of the data flow from the source node to the destination node. The route includes multiple dataflow links.
 
\subsection{TSN Switch Model}
The TSN switch is used for efficient data forwarding. We take Fig. \ref{TSN switch model}(b) as an example to illustrate the TSN switch model.

The data flow first enters the switch through the input port of the switch. Then, the input flow is processed by the forwarding engine, which determines the forwarding output port of the flow according to the route table. The forwarding engine assigns the flow to the corresponding queue dispatcher in TAS (Time-Aware Shaper). There are eight queues with different priorities (${Q}_7$ to ${Q}_0$) in the output port. ${Q}_7$ has the highest priority to store the TT (Time-Triggered) flows with strict deadline constraints; ${Q}_6$ and ${Q}_5$ have a lower priority than ${Q}_7$, and it stores the AVB\_A and AVB\_B flows, respectively; the remaining queues store the Best-Effort (BE) flows. TAS provides the GCL to manage the flow transmission by opening and closing the gates of the queues. When the gate of a queue is opened (denoted by ``o"), the data frame in the queue is allowed to transmit to the output port; when the gate operation is set to close (denoted by ``c"), the flow is prevented from being transmitted. As shown in Fig. \ref{TSN switch model}(b), only the gate of ${Q}_7$ (``o") is open, and the gates of other queues are closed (``c") at time ${t}_0$. Therefore, only the TT frame can be sent to the outport. 

The flow is sent out from the output port of the switch and enters the next node or the destination node. There are two scheduling modes named preemption (IEEE Qbv \cite{Qbv}) and non-preemption (IEEE Qbu \cite{Qbu}) in the TSN switch. We adopt the preemption mode to achieve the frame transmission, i.e., a higher-priority TT frame can interrupt the transmission of a lower-priority frame. 

\subsection{Flow Model}
\label{section:Message Model}
This paper focuses on the transmission of critical and non-critical flows. These two types of flows are modeled as separate sets: the critical flow set and the non-critical flow set. Consequently, the overall flow set in TSN is represented by the union of these two sets.

Each flow contains several attributes: flow type, transmission period, deadline, transmission duration time, send offset at the network node, source node, and destination node. The flow type indicates whether it is a critical or non-critical flow. The transmission period refers to the interval at which the flow is generated. The deadline specifies the time by which the flow must be delivered. The transmission duration time is the time required to complete the transmission of the flow. The send offset represents the scheduled transmission start time of the flow at a given network node; for example, the offset on a specific dataflow link between two nodes indicates when the flow is sent on that particular link. The source and destination nodes define the path endpoints for the flow.

For simplicity, this paper assumes that each flow consists of only one frame, and therefore the terms ``flow" and ``frame" are used interchangeably. The deadline of a flow does not exceed its period. In terms of frame size, the maximum payload of a TSN frame is 1500 bytes, with an additional transmission overhead of 42 bytes. We assume the transmission speed of the dataflow link is 100 Mbps. 

\section{Design of MCFS-2L}
This section presents the design of MCFS-2L. We first describe the overview of MCFS-2L. Then, we concentrate on the major technical components of MCFS-2L.

\begin{figure*}[ht]
	\centering
\includegraphics[width=1\textwidth]{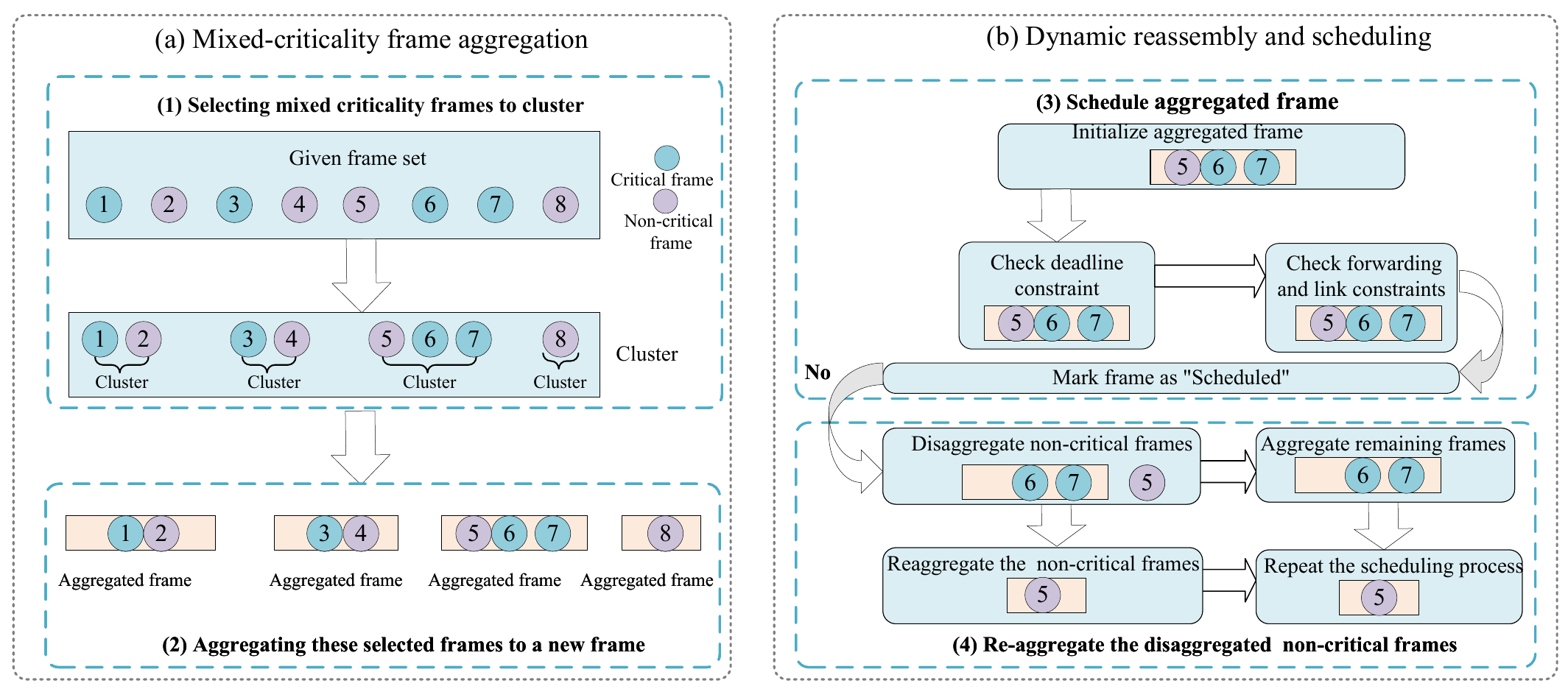}
	\caption{Overview of the MCFS-2L.}
	\label{Overview of the MCFS-2L}
\end{figure*}

\subsection{Overview}
\label{section:Overview}
Fig. \ref{Overview of the MCFS-2L} shows the workflow of MCFS-2L, which consists of two components, i.e., mixed-criticality frame aggregation and dynamic reassembly and scheduling. Through mixed-criticality frame aggregation, the critical and non-critical frames with the same source and destination nodes, as well as harmonic periods, are aggregated into the same frame. This method has the following advantages:

\textcolor{black}{\begin{itemize}
\item
\textbf{Fewer network frames.} By aggregating multiple small frames into a single larger frame, the total number of frames transmitted across the network is reduced, thereby lowering bandwidth overhead.
\item
\textbf{Fewer required transmission time windows.} Transmissions that previously needed individual time windows for each small frame can now be completed using a single window for the aggregated frame, greatly simplifying the scheduling process.
\item
\textbf{Improved bandwidth efficiency.} After aggregation, multiple data payloads share a single header, which avoids redundant transmission of protocol headers and enhances the effective utilization of available bandwidth.
\end{itemize}}

To further improve the acceptance ratio of frames, we adopt the dynamic reassembly and scheduling method. This method disaggregates the non-critical frames within the unscheduled frame. This scheduling method has the following advantages: 
\begin{itemize}
\item
\textbf{Improve the frame acceptance ratio}.  While aggregation enhances efficiency, it can also introduce a potential issue: if an aggregated frame contains an excessive amount of non-critical data, its size may exceed the maximum transmission limit of a given critical time window. In such cases, the entire aggregated frame becomes unschedulable. The dynamic disaggregation mechanism addresses this by separating the non-critical frames, ensuring that at least the critical frames can be scheduled, thereby enhancing the reliability of critical tasks.

\item
\textbf{Balance flexibility and reliability.} This approach embodies a combination of ``best-effort" and ``hard real-time guarantees". When bandwidth is abundant, non-critical frames are aggregated with critical frames for efficient transmission; however, during periods of bandwidth scarcity or contention, the convenience of non-critical frames is sacrificed to ensure the timely delivery of critical frames.

\item
\textbf{Reduce resource waste.} Dynamic disaggregation allows the system to maximally utilize the remaining bandwidth to schedule the disaggregated non-critical frames under existing resource constraints, thereby improving overall network resource utilization.

\end{itemize}

\subsection{Mixed-criticality Frame Aggregation}
\label{section:Mixed-criticality frame aggregation}
 In this phase, we aggregate the frames with mixed criticality levels to reduce the number of frames to be transmitted in TSN. In other words, we merge critical frames and non-critical frames that satisfy the aggregation constraints into one frame for transmission. The aggregation constraints are as follows.
 
\begin{itemize}
\item
\textbf{Source and destination nodes.} Criticality and non-criticality frames aggregated into the same frame must have the same source node and destination node. 
\item \textbf{Period}. The period of the new frame is the greatest common divisor of the periods of the criticality and non-criticality frames aggregated to this frame.

\item \textbf{Deadline}. The deadline of the new frame should be the minimum of the deadlines of critical and non-critical frames that are aggregated to this frame.

\item \textbf{Maximum transmission time.} The maximum frame size after aggregation should be less than 1542 bytes, that is, its maximum transmission time should be less than or equal to the maximum allowable transmission time.
\end{itemize}

Based on the aggregation constraints above, we will explain the detailed process of mixed-criticality frame aggregation, which consists of two steps: 1) selecting mixed-criticality frames to cluster; and 2) aggregating these selected frames to a new frame, as shown in Fig. \ref{Overview of the MCFS-2L}(a).

\textbf{(1) Selecting mixed criticality frames to cluster.} According to the aggregation constraints, we select the frames that meet the aggregation conditions. In other words, we select the frames with the same source and destination nodes and harmonic periods into the same cluster. For example, in Fig. \ref{Overview of the MCFS-2L}(a), the critical and non-critical frames meet the aggregation criteria and are grouped into the same cluster, where the curly braces indicate that the frames (critical or non-critical frames) that meet the aggregation constraints can be aggregated into the same cluster.

\textbf{(2) Aggregating these selected frames into a new frame.} After step 1, we aggregate the frames in the same cluster into a new frame. Note that the payload of the new frame after aggregation is less than or equal to 1500 bytes (i.e., the maximum transmission time). As shown in Fig. \ref{Overview of the MCFS-2L}(a), the critical and non-critical frames are aggregated into a new frame (the orange rectangle represents the new aggregated frame). For example, the new aggregated frame includes two critical frames (i.e., frames marked by ``6" and ``7") and one non-critical frame (i.e., frame marked by ``5"). In step 2, we only merge frames that meet the aggregation criteria into a new frame, while frames that cannot be aggregated remain unchanged.

\subsection{Dynamic Reassembly and Scheduling}
\label{section:Dynamic Reassembly and Scheduling}
After completing the aggregation of mixed-criticality frames, we then start scheduling frames. The scheduling of frames must satisfy the following constraints.

\begin{itemize}
\item
\textbf {Deadline constraint.}
  The deadline constraint means that the total latency of the frame from the source node to the destination node remains within its deadline.
\item \textbf {Forwarding constraint.}
Flow is sent in sequence through two consecutive dataflow links. This implies that the node must first receive flow before it is forwarded to the destination node. In other words, for any data flow, its transmission start time at the output port of the current switch must be greater than or equal to its completion time on the preceding link.
\item \textbf {Link constraint.}
The link constraint represents that one data link only deals with a data frame once. Otherwise, the transmission conflict will occur. In other words, to avoid transmission conflicts, no two TSN frames can be sent to the same data link at the same time.
\end{itemize}

According to the scheduling constraints above, we start scheduling the frames based on their priorities. To improve the acceptance ratio, we propose a dynamic reassembly and scheduling method to deal with the unschedulable frames, as shown in Fig. \ref{Overview of the MCFS-2L}(b). The detailed process is outlined as follows.

\textbf{(3) Schedule aggregated frame.} The frame with a smaller deadline is assigned a higher priority. A frame is considered schedulable if it can complete transmission within its deadline; otherwise, it is considered as an un-schedulable frame. If all constraints are satisfied, the frame is successfully scheduled at the offset. Otherwise, if either the forwarding constraint or link constraint is violated, the offset is incremented by one unit, and the verification process repeats.  

\textbf{(4) Re-aggregate the disaggregated non-critical frames.} For the unscheduled frame, this method identifies the non-critical frames within the original frame. Then, sequentially extract each non-critical frame and place it into a specified set. Meanwhile, we aggregate the remaining frames in the original frame, which continues to be merged into a new frame for rescheduling. For example, the non-critical frame marked by ``5"  in Fig. \ref{Overview of the MCFS-2L}(b) is considered as a new frame for rescheduling.
Finally, reschedule the new frame after aggregation. The non-critical frames extracted are also aggregated into a new frame for scheduling.

\section{Case study}
In the case study, we show the effectiveness of our MCFS-2L in acceptance ratio, bandwidth utilization, and execution time.

\subsection{Experimental Setup}

\textbf{Dataset.} We use a real-world dataset from General Motors \cite{zhou2021reliability}, which is composed of multiple flows from applications such as active safety, engine control, automated driving, and surround view. The sending domain of each flow is randomly selected from the ADAS and intelligent cockpit domains, and the intra-domain network adopts the TSN. The destination domain of flows is the vehicle control domain. The flow periods are randomly generated between 20 and 100 ms, with deadlines ranging from 200 $\mu$s to 800 $\mu$s. In this study, the link transmission speed is set to 100 Mbps. The TSN frame payload ranges from 100 to 1500 bytes,  and each frame includes an additional 42-byte overhead. 

\textbf{Baselines.} To evaluate the performance of MCFS-2L, we compare with two representative baselines, i.e., NWTT \cite{durr2016no} and R-NWTT \cite{lin2022rethinking}. NWTT sets the initial scheduling time of flows to the earliest possible start time. R-NWTT is a scheduling algorithm that randomly determines a schedulable start time for each flow within the range of 0 to its relative deadline.

\subsection{Evaluation Metrics}
\textbf{Acceptance ratio of critical frames.} The performance metric used in the experiment is the acceptance ratio of critical frames. This ratio is defined as the number of critical frames that successfully meet their deadlines divided by the total number of critical frames transmitted within a given time interval. A higher acceptance ratio of critical frames indicates a better scheduling performance of MCFS-2L.

\textbf{Acceptance ratio of non-critical frames.} This ratio is defined as the number of non-critical frames that successfully meet their deadlines divided by the total number of non-critical frames transmitted within a given time interval. A higher acceptance ratio of non-critical frames indicates better scheduling performance of MCFS-2L, resulting in improved Quality of Service (QoS) \cite{zhou2025user}.

\textbf{Bandwidth utilization.} Bandwidth utilization refers to the percentage of time occupied by the transmission of all schedulable frames from their source nodes to their destination nodes, relative to the total duration of the hyperperiod for all frames to be transmitted. Specifically, the total bandwidth utilization is calculated by summing the transmission times of all frames and then dividing that sum by the hyperperiod.

\textbf{Execution time.} This metric measures the amount of time required for the MCFS-2L to complete its tasks under varying conditions.

\subsection{Performance Evaluation}
\subsubsection{\textcolor{black}{Acceptance Ratio of Critical Frames}}
\textcolor{black}{We first evaluate the acceptance ratio of critical frames comparison under the varying TSN frame scales in the preemption mode. The number represents the total count of TSN frames, including critical and non-critical frames in the experiment. }

\begin{figure}[t]
    \centering
    % 再往左一点点：-13.0mm → -13.2mm
    \hspace{-13.185mm}
    \begin{subfigure}[b]{0.51\textwidth}
        \centering
        \includegraphics[width=0.85\textwidth, trim=18 0 0 0, clip]{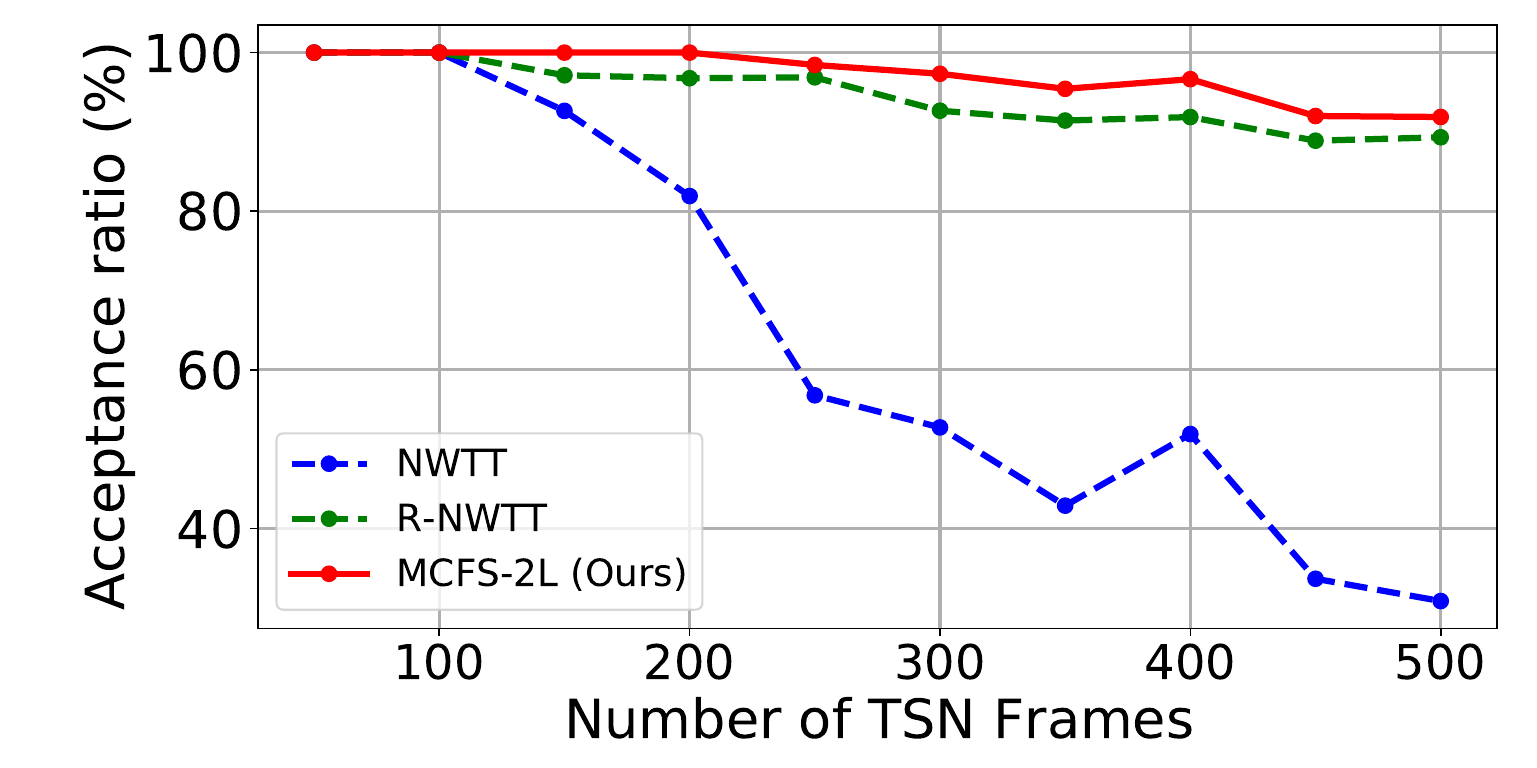}
        \caption{Acceptance ratio of critical frames.}
        \label{fig:acceptance_ratio_of_critical_frames}
    \end{subfigure}
    
    \vspace{6mm}
    
    \hspace{-13.185mm}
    \begin{subfigure}[b]{0.51\textwidth}
        \centering
        \includegraphics[width=0.85\textwidth, trim=10 0 0 0, clip]{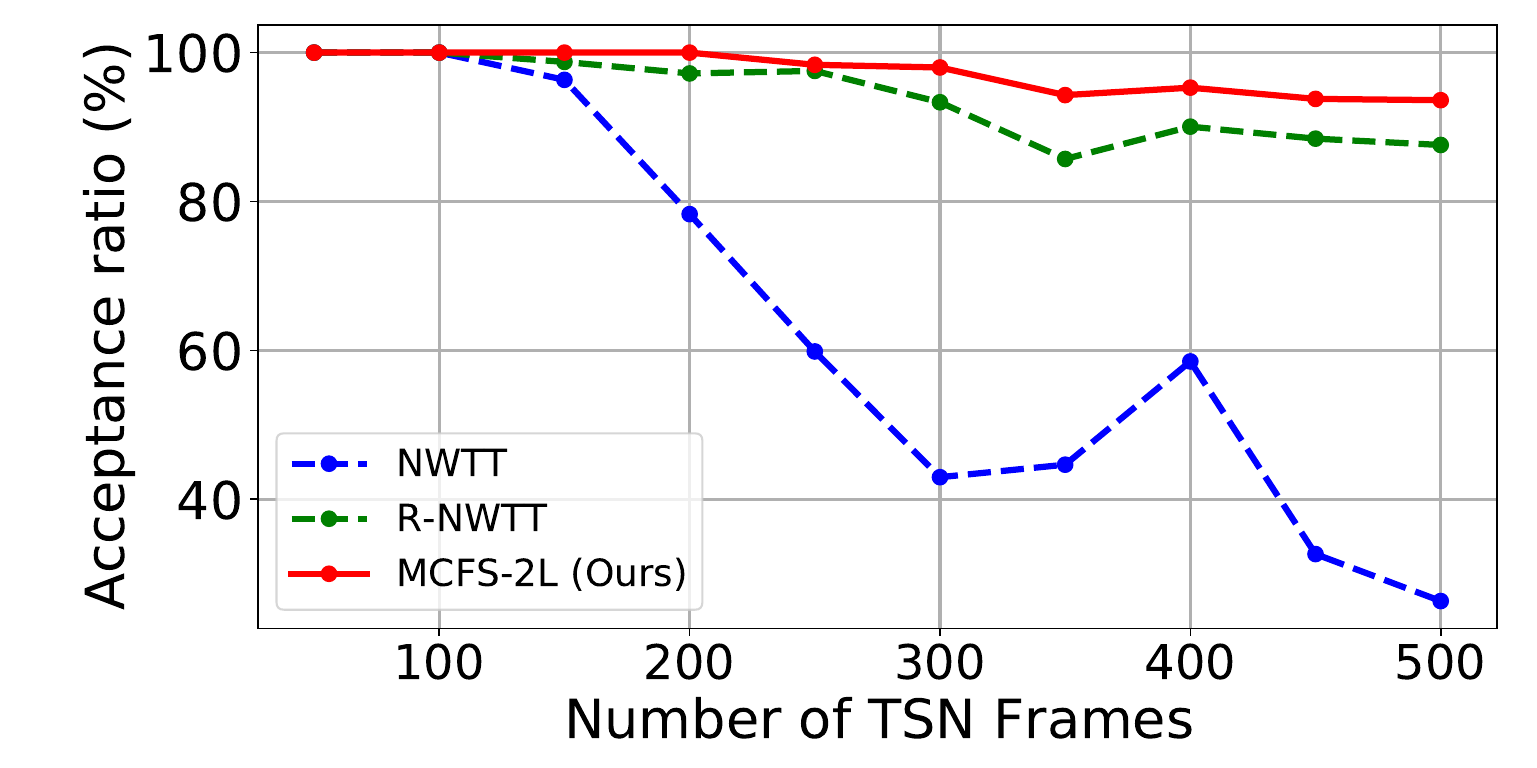}
        \caption{Acceptance ratio of non-critical frames.}
        \label{fig:acceptance_ratio_of_noncritical_frames}
    \end{subfigure}
    
    \caption{\textcolor{black}{Acceptance ratio of critical frames and non-critical frames under different numbers of TSN frames.}}
    \label{fig:acceptance_ratio_combined}
\end{figure}
Fig. \ref{fig:acceptance_ratio_combined}(a) shows the results of the acceptance ratio of critical frames under varying numbers of TSN frames. We observe that the acceptance ratio of critical frames of NWTT, R-NWTT, and MCFS-2L remains at 100\% when the number of TSN frames is relatively small (e.g., 50 or 100 TSN frames). However, as the number of TSN frames increases, the acceptance ratio of critical frames for both methods shows a downward trend, and individual data fluctuates due to the random generation of the attributes of TSN frames. We further observe that the acceptance ratio of critical frames of MCFS-2L is generally higher than that of NWTT and R-NWTT under the same number of TSN frames. This is because MCFS-2L reduces the number of transmitted frames through frame aggregation on the one hand, and improves the acceptance ratio of critical frames of the unscheduled frames through dynamic reassembly and scheduling on the other hand.

\subsubsection{\textcolor{black}{Acceptance Ratio of Non-Critical Frames}} 
\textcolor{black}{Then, we compare the acceptance ratio of non-critical frames with varying TSN frame scales.} Fig. \ref{fig:acceptance_ratio_combined}(b) presents the experimental results of the acceptance ratio of non-critical frames of different methods. We observe that the acceptance ratio of non-critical frames of MCFS-2L is consistently higher than that of NWTT and R-NWTT under the same number of TSN frames. For example, the acceptance ratio of non-critical frames of MCFS-2L is higher than that of R-NWTT by 6.02\% when the number of TSN frames is 500. This is because the aggregation of some non-critical frames reduces the number of non-critical frames to be transmitted. As the number of TSN frames increases, the acceptance ratio of non-critical frames of NWTT, R-NWTT, and MCFS-2L both decreases.

\subsubsection{\textcolor{black}{Bandwidth Utilization}}
\begin{figure}[t]
    \centering
% \hspace{18mm}  % 向左移动图像
\includegraphics[width=0.44\textwidth]{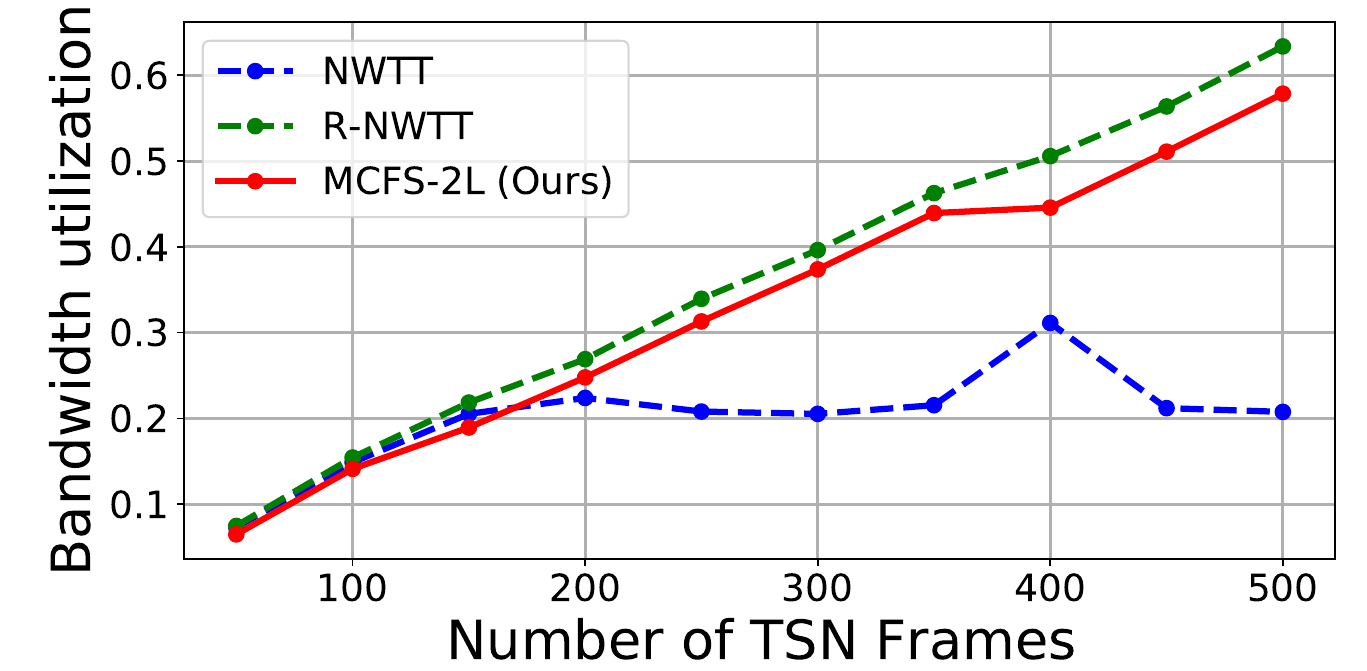}
    \caption{\textcolor{black}{Bandwidth utilization under different numbers of TSN frames.}}
    \label{fig:BW}
\end{figure}

\textcolor{black}{We also conduct a comparison of the bandwidth utilization, as shown in Fig. \ref{fig:BW}.} We observe that the bandwidth utilization of MCFS-2L is lower than that of NWTT and R-NWTT for the same number of TSN frames. For instance, when the number of TSN frames is 50 or 100, the acceptance ratio of critical frames and the acceptance ratio of non-critical frames are both 100\% among the three methods, whereas the bandwidth utilization of MCFS-2L is lower than that of NWTT and R-NWTT. As the number of TSN frames increases, the acceptance ratio of critical frames and the acceptance ratio of non-critical frames of MCFS-2L are higher than those of NWTT and R-NWTT, but the bandwidth utilization of MCFS-2L is still lower than that of R-NWTT. At this time, NWTT has the lowest bandwidth utilization because it has the lowest acceptance ratio of critical frames and the acceptance ratio of non-critical frames. Therefore, we conclude that MCFS-2L achieves a lower bandwidth utilization than R-NWTT. 
 
\subsubsection{\textcolor{black}{Execution Time}}
\begin{figure}[t]
	\centering
     \hspace{-4mm}  % 向左移动图像
\includegraphics[width=0.47\textwidth]{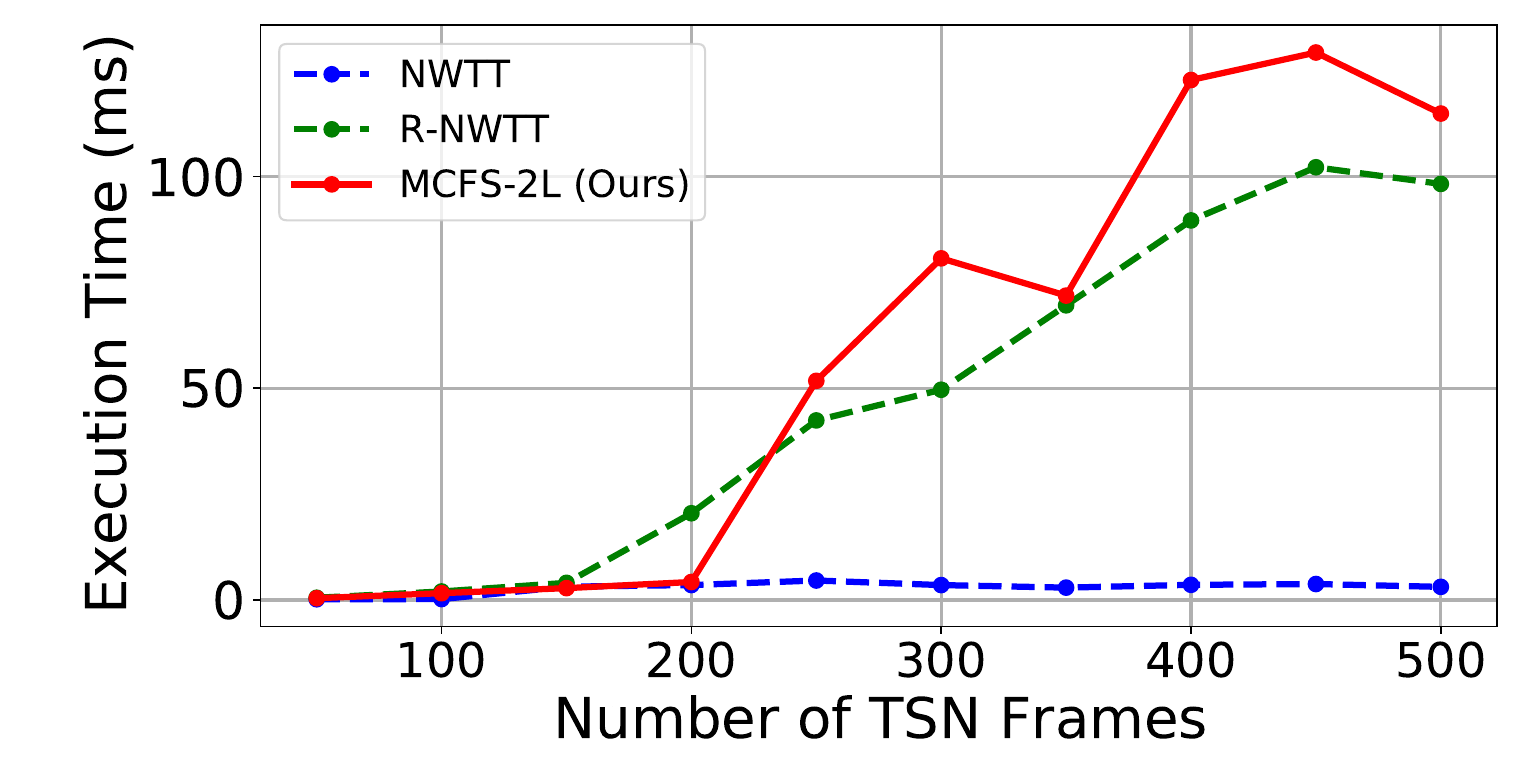}
	\caption{Execution time comparison.}
	\label{Execution time comparison}
\end{figure}

\textcolor{black}{Finally, we compare the execution time. As shown in Fig. \ref{Execution time comparison}, when the number of the TSN frames is small, the execution time of MCFS-2L is lower than that of R-NWTT.} However, when the number of TSN frames reaches or exceeds about 250, the execution time of MCFS-2L is higher than that of R-NWTT. The reason is as follows. As the number of TSN frames increases, more unscheduled frames need to be processed. MCFS-2L splits more non-critical frames from the aggregated frames to increase the acceptance ratio, which leads to an increase in time complexity. NWTT has the lowest acceptance ratio, thereby resulting in the lowest execution time.

\section{Conclusion}
\label{section:Conclusion}

\textcolor{black}{In this paper, we studied a mixed-criticality flow scheduling scheme, MCFS-2L, to address the challenges of low delay and limited bandwidth in flow transmission. MCFS-2L introduces a mixed-criticality frame aggregation strategy that aggregates critical and non-critical frames into larger frames to reduce the number of transmitted frames. It then employs an effective frame scheduling algorithm to schedule the aggregated frames and improve the acceptance ratio of critical flows. Experimental results demonstrate that MCFS-2L significantly increases the acceptance ratio of both critical and non-critical frames and improves bandwidth utilization compared with existing state-of-the-art methods. MCFS-2L opens a new research direction in TSN-based mixed-criticality systems. Future work will extend MCFS-2L to different network architectures and integrate adaptive mechanisms that can dynamically respond to varying traffic patterns.}

\addtolength{\topmargin}{-0.025in}
\ifCLASSOPTIONcompsoc
\else
\fi

\bibliography{HCO3.bib}
\bibliographystyle{IEEEtran}
\end{document}